\documentclass[prd,preprintnumbers,nofootinbib,showpacs,12pt]{revtex4-1}

\usepackage{bm}
\usepackage{amsmath,amsbsy,graphicx,amsfonts,mathrsfs,amssymb}
\usepackage{latexsym}
\usepackage{graphicx,picinpar,epsfig}
\usepackage{pdfsync}
\usepackage{float}
\usepackage{color}
\usepackage[hyperindex,colorlinks,citecolor=black,linkcolor=black,CJKbookmarks=true]{hyperref}
\usepackage{calc}

\newcommand{\p}{\partial}
\newcommand{\del}{\nabla}
\newcommand{\tr}{\text{tr}}

\newcommand{\dt}{\text{det}}

\begin{document}

\title{Fluid/gravity correspondence: A nonconformal realization in compactified D4 branes}

\author{Chao Wu$^{1}$}
\email{wuchao@ihep.ac.cn}
\author{Yidian Chen$^{1}$}
\email{chenyd@ihep.ac.cn}
\author{Mei Huang$^{1,2}$}
\email{huangm@ihep.ac.cn}
\affiliation{$^1$ Institute of High Energy Physics, Chinese Academy of Sciences, Beijing 100049, China}
\affiliation{$^2$ Theoretical Physics Center for Science Facilities, Chinese Academy of Sciences, Beijing 100049, China}
\date{\today}

\begin{abstract}
  We develop the framework of boundary derivative expansion (BDE) formalism of fluid/gravity correspondence in compactified D4-brane system, which is a nonconformal background used in top-down holographic QCD models. Such models contain the D4-D6 model and the Sakai-Sugimoto (SS) model, with the background of the compactified black D4 branes under the near horizon limit. By using the dimensional reduction technique, we derive a 5D Einstein gravity minimally coupled with 3 scalar fields from the 10D D4-brane background. Following the BDE formalism of fluid/gravity correspondence in the conformal background, we directly derive all the first order transport coefficients for nonconformal gluonic matter. The results of the ratio of the bulk to shear viscosity and the sound speed agree with those obtained from the Green-Kubo method. This agreement guarantees the validity of the BDE formalism of fluid/gravity duality in the nonconformal D-brane background, which can be used to calculate the second order transport coefficients in nonconformal background.
\end{abstract}

\pacs{11.25.Tq}
\maketitle

\section{Introduction}

Studying Quantum chromodynamics (QCD) phase transition and properties of hot/dense quark matter at high temperature and baryon density is one of the most important topics of high energy nuclear physics. The Relativistic Heavy Ion collider (RHIC) and the Large Hadron Collider (LHC) provide the opportunity to investigate properties of nuclear matter at high temperature and small baryon density. It is now believed that the system created at RHIC/LHC is a strongly coupled quark-gluon plasma (sQGP) and behaves like a nearly ``perfect" fluid \cite{RHIC-EXP,RHIC-THEO}. One crucial quantity is the shear viscosity over entropy density $\eta/s$, which is required to be very small to fit the elliptic flow at RHIC/LHC. The result from AdS/CFT correspondence gives the lower bound of $\eta/s= \frac{1}{4\pi}$ \cite{KSS_bound,Kovtun:2004de}, which is very close to the value used to fit the elliptic flow $v_2$ \cite{Song:2008hj,Hydro, Hydro-Teaney}.

The  anti-de Sitter/conformal field theory (AdS/CFT) duality \cite{Maldacena:AdS/CFT,GKP:AdS/CFT,Witten:AdS/CFT} is  discovered through pioneering works on the near horizon structure of black branes (or black holes) \cite{near_horizon_struct1,near_horizon_struct2,near_horizon_struct3,near_horizon_struct4,near_horizon_struct5,near_horizon_struct6,
near_horizon_struct7} and the scattering process of branes and bulk probe fields \cite{brane_absorption1,brane_absorption2,brane_absorption3,brane_absorption4}. It is generalized to nonconformal brane background in the near horizon limit in ref. \cite{Itzhaki4:gauge/gravity}, which is called the gauge/gravity duality nowadays. The gravity/gauge duality or AdS/CFT correspondence provides a revolutionary method to tackle the problem of strongly coupled gauge theories. It has been widely used to investigate QCD phenomenology, e.g.
glueballs \cite{Earliest_Glueball1,Earliest_Glueball2,Earliest_Glueball3}, hadron spectra \cite{D3D7-I,D3D7-II,D3D7-III,D4D6,D4D8}, deconfinement phase transition \cite{D3D7-II,D3D7-III,D4D6,phase_transit_SS} and transport properties \cite{Policastro:eta_N=4_SYM_plasma}.

The shear viscosity in AdS/CFT was firstly calculated in Ref. \cite{Policastro:eta_N=4_SYM_plasma} through relations between Green-Kubo formula\footnote{which is first proposed by Ryogo Kubo \cite{Kubo_formulae} in statistical mechanics and recast into field theory formalism by Akio Hosaya et.al. in \cite{Hosaya:Kubo_form_particle_phys}.} and the absorption cross section of gravitons \cite{brane_absorption1,brane_absorption2,brane_absorption3,brane_absorption4}.
Studies on the near equilibrium QGP from AdS/CFT duality in \cite{Son:real_time_AdS/CFT} gives a recipe of extracting two-points real-time thermal correlators via classical bulk action. Following \cite{Son:real_time_AdS/CFT}, the authors of \cite{Policastro:AdS/CFT_hydro1,Policastro:AdS/CFT_hydro2} calculated the first order transport coefficients in near extremal D3 brane background and found that, in the long-distance and low-frequence limit, these correlators turn into hydrodynamical forms. Second order transport coefficients of this system were calculated in \cite{Baier:AdS/CFT_hydro_2order}. The framework that investigates transport properties of a fluid via its corresponding gravity is called the fluid/gravity correspondence, and the most notable feature in above works \cite{Policastro:AdS/CFT_hydro1,Policastro:AdS/CFT_hydro2,Baier:AdS/CFT_hydro_2order} is the use of Green-Kubo formula, thus one may call it the Green-Kubo formalism of fluid/gravity correspondence.

While the Green-Kubo formalism becomes popular in extracting transport properties of liquid-like plasma\footnote{the literature on this topic can be found in the references of \cite{Son;1order_hydro_review}.}, another systematical and powerful formalism---the boundary derivative expansion (BDE) formalism \cite{Bhattacharyya:fluid/gravity,Bhattacharyya:Js} has been developed. The most remarkable feature for this formalism is the use of boundary dependent boost parameters for the bulk metric in the (in-going) Eddington-Finkelstein coordinate. Expansions are implemented with respect to boundary derivatives of the boost parameters and all the dissipative terms of boundary fluid are metric perturbations (in the large $r$ limit) solved from Einstein equation.
The first example of the BDE formalism of fluid/gravity correspondence is a duality of $AdS_5$ black hole in the bulk to finite temperature conformal $\mathcal N=4$ SYM plasma in the boundary \cite{Bhattacharyya:fluid/gravity}, where the transport coefficients were calculated to second order. The BDE formalism was applied in several other models with $AdS_5$ black hole background: 1. The $AdS_5$-dilaton model \cite{AdS5_dilaton_hydro}, where the gravity side is an $AdS_5$ black hole plus a boundary dependent dilaton field, while in the boundary is a fluid with forcing terms. 2. The charged $AdS_5$ black hole model \cite{Banerjee:AdS5Q_hydro,Erdmenger:AdS5Q_hydro}, where the bulk is a charged $AdS_5$ black hole, and the fluid on the boundary has a chemical potential. The addition of the Chern-Simons term for the $U(1)$ gauge field cause the appearance of vorticity in the first order dissipative expansion of R-charge current.

The development of fluid/gravity correspondence interweaves with the studies on hydrodynamical modes on the world-volume of the blackfold \cite{blackfold1,blackfold2,blackfold3,blackfold4}, which opens a window to extract the dynamical information on the world-volume of black branes in flat spacetime. The most obvious difference of this kind of research from the fluid/gravity correspondence is the need of Dirichlet boundary condition on a finite cut-off surface. In \cite{Emparan:blackfold_hydro}, the effective hydrodynamics on a $p+n+2$ dimensional ``rigid wall" located at $r=R$ in a $D=p+n+3$ dimensional spacetime is studied, where $p$ and $n+1$ are the number of (spatial) dimension of the brane and the sphere, respectively. Based on this, \cite{Emparan:hydro_black_D3} studies the effective hydrodynamics on the world-volume of black D3 brane to first order, both the thermodynamical and the viscous quantities depend on the location of the cavity and the horizon, however $\eta/s$ of this model is still $\frac1{4\pi}$. Ref. \cite{Erdmenger:hydro_spin_D3} investigates the effective hydrodynamics of rotating black D3 branes. The common feature of \cite{Emparan:hydro_black_D3,Erdmenger:hydro_spin_D3} is the use of dimensional reduction, which transforms the effects of transverse directions into massless fields on the longitudinal directions in which the branes lie. Thus makes us focus on the worldvolume theory.

The Green-Kubo formalism and BDE formalism of  fluid/gravity correspondence provide powerful tools for us to study the liquid-like QGP (see e.g., \cite{LiuHong:hot_QCD,Iancu:hearvy_ion} for a phenomenological review on heavy ion collisions for theorists). Generally speaking, QGP is a liquid-like plasma with small shear viscosity, and can be described by relativistic hydrodynamics quite well. Results from lattice show that QGP exhibits nonconformal properties especially around the critical temperature $T_c$, e.g. the shear viscosity over entropy density ratio has a minimum around $T_c$, and the bulk viscosity over entropy density shows a peak around $T_c$ \cite{LAT-etas,LAT-xis-KT,LAT-xis-Meyer,correlation-Karsch}. This behavior has been described in bottom-up holographic QCD models \cite{bottomup-gubser,bottomup-kampfer,bottomup-noronha,bottomup-huangli}. However, current studies using fluid/gravity duality from top-down method are mostly on AdS background, whose dual fluid is of course conformal, thus may only reflect the properties of QGP at the conformal regime, i.e., above 2$T_c$. However, when we are at the nonconformal zone around $T_c$, AdS gravity may no longer be proper for a nonconformal gauge theory.

A natural choice in top-down holographic way to tackle the strongly coupled nonconformal plasma is to build models using the nonconformal D-branes. Such studies include e.g. \cite{David:D1-brane_hydro} for D1-brane and \cite{Mas:Dp_hydro_GreenKubo} for Dp-branes with $p\geq2$, where the Green-Kubo method is used. There is another interesting work on this topic which can handle more cases including $p=0,1$ and the fundamental strings in type II string theory (but $p=5$ excluded): the Ref. \cite{Skenderis:Dp-brane_hydro}, where the BDE formalism in Fefferman-Graham coordinates developed in \cite{Gupta:BDE_FeffermanGraham} is used. Besides the Dp-brane backgrounds like in Refs. \cite{Mas:Dp_hydro_GreenKubo,Skenderis:Dp-brane_hydro}, one may also use the compactified D-brane backgrounds, e.g. the compactified D4-brane. The compactified D4-brane is the background of D4-D6 model \cite{D4D6} and the Sakai-Sugimoto (SS) model \cite{D4D8}, which are two nonconformal holographic QCD models from top-down. The background of these two models is the compactified D4 black branes under the near horizon limit. There are some previous studies on the transport properties of this background. The sound speed and bulk to shear ratio was calculated in \cite{Benincasa:hydro_SS}, the shear to entropy density ratio was argued in this reference to be $1/{4\pi}$ by showing that the SS model background is in the class of \cite{Buchel:universality_shear}. Using the null horizon focusing equation,  Ref. \cite{Eling:novel_formla_bulk_viscs} also calculated the ratio of the bulk viscosity to the shear viscosity.

Based on the above review of the relevant literatures, one can see that there is a lack of parallel formulation with \cite{Bhattacharyya:fluid/gravity}. This motivates us to develop the framework of BDE formalism of fluid/gravity duality for nonconformal gauge theory plasma. In this paper, by using the BDE formalism of fluid/gravity correspondence, we calculate the first order transport coefficients of the nonconformal QGP under the quenched limit on the boundary of both the D4-D6 and SS model's background. Our results are consistent with former studies by other methods. The previous results together with ours reveal that the plasma of the D4-D6 and SS model is nonconformal with a small bulk viscosity and saturates the KSS bound \cite{KSS_bound,Kovtun:2004de}, and this agreement guarantees the validity of the BDE formalism of fluid/gravity correspondence for nonconformal D-brane backgrounds with more than one submanifolds reduced. This work can been seen as a nonconformal counterpart that is parallel with Bhattacharyya et al.'s $AdS_5$ construction in fluid/gravity correspondence \cite{Bhattacharyya:fluid/gravity}.

This paper is organized as follows: After introduction, we will firstly give the preliminaries from 10D compactified black D4 brane background to a 5 dimensional one in section 2 in order to make connection with the recipe of fluid/gravity correspondence. Then, in section 3, we will solve all the first order perturbative ansatz and get the metric which perturbatively solves the Einstein equation to the first order. By making use of this solution, we calculate the boundary stress tensor for the QGP which corresponds to the bulk of SS model in section 4 and analyze its transport properties. We give the discussion and outlook in section 5.

\section{The Setup}

In this section, following \cite{Benincasa:hydro_SS}, we will show how to derive the action and classical background of the D4-D6 and SS model into 5D form through dimensional reduction as in \cite{Emparan:hydro_black_D3,Erdmenger:hydro_spin_D3}. The purpose of doing this is to make connection with \cite{Bhattacharyya:fluid/gravity}, more details can be found in Appendix \ref{Apendix}.

The D4 brane action of type IIA theory in Einstein frame is given as
\begin{align}
  S=\frac1{2\kappa_{10}^2}\int d^{10}x\sqrt{-G}\left[ R^{(10)}-\frac12(^{10}\del\phi)^2-\frac{g_s^2}{2\cdot4!}e^{\frac{\phi}2}F_4^2 \right],
  \label{eq:bulk action}
\end{align}
where $2\kappa_{10}^2=(2\pi)^7g_s^2l_s^8$ is the 10D gravitational coupling and $^{10}\del$ stands for 10D nabla. $G$ is the determinant of the following diagonal 10D metric tensor:
\begin{align}\label{reduction ansatz}
  ds^2=e^{2\alpha_1A}g_{MN}dx^Mdx^N+e^{2\alpha_2A}\left( e^{2\beta_1B}dy^2
         +e^{2\beta_2B}\gamma_{ab}d\theta^ad\theta^b \right),
\end{align}
where $g_{MN}$, $A$ and $B$ only depend on $x^M$, the coordinates of first 5 dimensions, and $\gamma_{ab}$ with $a,b=1,2,3,4$ is the metric on the $S^4$. $\alpha_{1,2}$ and $\beta_{1,2}$ are four parameters whose value will be clear in the following context. The explicit form of dilaton and RR field are given in (\ref{Einstein frame Black D4}). It should be noticed here that $y$ is also a compact dimension and we will integrate out both $y$ and the 4-sphere to get a 5D effective theory.

From Eq.(\ref{reduction ansatz}), we have $\sqrt{-G}=e^{5(\alpha_1+\alpha_2)A+(\beta_1+4\beta_2)B}\sqrt{-g}\sqrt\gamma$ with $\gamma=\dt\gamma_{ab}$ the determinant of the metric on unit 4 sphere. During the reduction process, we have used the following relation:
\begin{align}
  S \sim \int d^{10}x\sqrt{-G}(R^{(10)}+\cdots) = \int d^5x \sqrt{-g}e^{(3\alpha_1+5\alpha_2)A+(\beta_1+4\beta_2)B}(R+\cdots).
\end{align}
To avoid the appearance of non-minimal coupling of the gravity with the scalar field in the reduced theory, one should set
\begin{align}
  \alpha_1=-\frac53,~~\alpha_2=1,~~\beta_1=4,~~\beta_2=-1,
  \label{eq:reduction parameter}
\end{align}
so Eq.(\ref{reduction ansatz}) becomes
\begin{align}\label{eq:reduct ansatz concrete No.}
  ds^2=e^{-\frac{10}3A}g_{MN}dx^Mdx^N+e^{2A+8B}dy^2+e^{2A-2B}d\Omega_4^2.
\end{align}
From Eq.(\ref{reduction of Ricci Scalar}), the 10D Ricci scalar has the form of:
\begin{align}\label{eq:Ricci scalar 10 to 5}
  R^{(10)}=e^{\frac{10}3A}\left[R+\frac{10}3\del^2A-\frac{40}3(\p A)^2-20(\p B)^2\right]+12e^{-2A+2B}.
\end{align}
During the reduction process, we have
\begin{align}
  \sqrt{-G}&=\sqrt{-g}\sqrt\gamma e^{-\frac{10}3A},\\
  \sqrt{-G}R^{(10)}&=\sqrt{-g}\sqrt\gamma\left(R+\frac{10}3\del^2A-\frac{40}3(\p A)^2-20(\p B)^2+12e^{-\frac{16}3A+2B}\right), \\
  \sqrt{-G}(^{10}\del\phi)^2&=\sqrt{-g}\sqrt\gamma e^{-\frac{10}3A}G^{MN}\p_M\phi\p_N\phi=\sqrt{-g}\sqrt\gamma(\p\phi)^2, \\
  \sqrt{-G}\frac{g_s^2}{2\cdot4!}e^{\frac{\phi}2}F_4^2&=\sqrt{-g}\sqrt\gamma\frac{Q_4^2}2e^{\frac\phi2-\frac{34}3A+8B}.
\end{align}
Therefore the D4 brane action Eq.(\ref{eq:bulk action}) is reduced to its 5D form and takes the form as
\begin{align}\label{eq:5D bulk act}
  S&=\frac1{2\kappa_5^2}\int d^5x\sqrt{-g}\left[R-\frac12(\p\phi)^2-\frac{40}3(\p A)^2-20(\p B)^2-V(\phi,A,B)\right],\cr
  V&(\phi,A,B)=\frac{Q_4^2}2e^{\frac\phi2-\frac{34}3A+8B}-12e^{-\frac{16}3A+2B},
\end{align}
where $\kappa_5$ is the 5D surface gravity with the following definition:
\begin{align}
  \frac1{2\kappa_5^2}\equiv{V_1\Omega_4\over2\kappa_{10}^2},
\end{align}
with $V_1=\int dy$ the volume of the compact circle. The system turns into a 5D Einstein gravity minimally coupled with 3 scalars $\phi,A,B$,
and $V(\phi,A,B)$ is the scalar potential.
The EOMs for this reduced system are:
\begin{align}
  E_{MN}&-T_{MN}=0,\label{eq:5D Ein eq} \\
  \del^2\phi&-\frac{Q_4^2}4e^{\frac\phi2-\frac{34}3A+8B}=0, \label{eq:5D phi eq} \\
  \del^2A&+\frac{17Q_4^2}{80}e^{\frac\phi2-\frac{34}3A+8B}-\frac{12}5e^{-\frac{16}3A+2B}=0,\label{eq:5D A eq} \\
  \del^2B&-\frac{Q_4^2}{10}e^{\frac\phi2-\frac{34}3A+8B}+\frac35e^{-\frac{16}3A+2B}=0, \label{eq:5D B eq}
\end{align}
where
\begin{align}
  E_{MN}\equiv R_{MN}-\frac12g_{MN}R
\end{align}
is the Einstein tensor in the 5D spacetime, and
\begin{align}
  T_{MN}&\equiv \frac12\left(\p_M\phi\p_N\phi-\frac12g_{MN}(\p\phi)^2\right)+\frac{40}3\left(\p_M A\p_N A-\frac12g_{MN}(\p A)^2\right)\cr
  &+20\left(\p_M B\p_N B-\frac12g_{MN}(\p B)^2\right)-\frac12g_{MN}V,
\end{align}
which can be viewed as the energy-momentum tensor in the bulk.

The classical solution for black D4 brane in Einstein frame reads
\begin{align}  \label{Einstein frame Black D4}
  ds^2&=H_4^{-\frac38}(-f(r)dt^2+d\vec x^2)+H_4^{\frac58}\frac{dr^2}{f(r)}+H_4^{-\frac38}dy^2+H_4^{\frac58}r^2d\Omega_4^2, \cr
  e^\phi&=e^{\Phi-\Phi_0}=H_4^{-\frac14},~~~F_4=g_s^{-1}Q_4\epsilon_4,~~~H_4=1+\frac{r_{Q4}^3}{r^3},~~~f(r)=1-\frac{r_H^3}{r^3},
\end{align}
where $g_s=e^{\Phi_0}$ and $Q_4=(2\pi l_s)^3g_sN_c/\Omega_4$.\footnote{The normalization condition for $Q_4$ here is $2\kappa^2\mu_4N_c=\int_{S^4}F_4$, where $2\kappa^2=2\kappa_{10}^2g_s^{-2}$ and $\mu_4=((2\pi)^4l_s^5)^{-1}$ is the D4 brane charge.} Note that we write one of the directions that the D4 brane lies (denoted by $y$) separatedly from the other 3 directions (denoted by \{$\vec x$\}) in order to  compare with (\ref{eq:reduct ansatz concrete No.}). Under the near horizon limit, the above metric becomes
\begin{align}
  ds^2&=\left(\frac rL\right)^\frac98(-f(r)dt^2+d\vec x^2)+\left(\frac Lr\right)^\frac{15}8\frac{dr^2}{f(r)}+\left(\frac rL\right)^\frac98dy^2
  +L^\frac{15}8r^\frac18d\Omega_4^2, \label{eq:n.h.l. D4 metric}\\
  e^\phi&=\left(\frac rL\right)^\frac34.\label{eq:NHL_dilaton}
\end{align}
where $L^3=Q_4/3=\pi g_sN_cl_s^3$. The above metric differs from the D4-D6 model and the SS model for the interchange of $t$ with $y$. Besides, it is in the Einstein frame, not string frame. Comparing Eq.(\ref{eq:reduct ansatz concrete No.}) with Eq.(\ref{eq:n.h.l. D4 metric}), we have
\begin{align}\label{eq:0th reduc soluti}
  e^A=L^{\frac{51}{80}}r^\frac{13}{80},~~~~e^B=L^{-\frac3{10}}r^{\frac1{10}},
\end{align}
and the reduced 5D metric is
\begin{align}\label{eq:5d bulk diagnal metrc}
  ds^2=Lr^\frac53(-f(r)dt^2+d\vec x^2)+\frac{L^4}{r^\frac43f(r)}dr^2.
\end{align}
From its Ricci scalar $R\sim-\frac5{6r^{11/3}}(14r^3+r_H^3)$, when $r\to0$, $R$ will become minus infinity so $r=0$ is the curvature singularity and away from that point the above metric will always be regular, thus we will only focus on the regime of $r>0$ from now on. At the boundary $r\to\infty$, $R\to0$ so Eq.(\ref{eq:5d bulk diagnal metrc}) is asymptotically flat, which is not obvious for the appearance of $r^{5/3}$ in the first 4 dimensions.

We turn to the in-going Eddington-Finkelstein coordinate by making the following transformation as $dt=dv-\frac{L^{3/2}}{r^{3/2}f(r)}dr$, then the above metric becomes
\begin{align}
  ds^2=Lr^\frac53(-f(r)dv^2+d\vec x^2)+2L^{\frac52}r^\frac16dvdr.
\end{align}
$r=0$ is still the curvature singularity of this 5D metric but everywhere away from that is regular. Since we have already lost the track of dimensions in the process of dimensional reduction Eq.(\ref{reduction ansatz}), keeping $L$ explicit will be insignificant, so from now on we set $L=1$, which means $Q_4=3$. After a boost of coordinate: $dv=-u_\mu dx^\mu,~dx^i=P^i_{~\mu}dx^\mu$, where $P_{\mu\nu}=\eta_{\mu\nu}+u_\mu u_\nu$, we have
\begin{align}\label{eq:boosted 5d bulk}
  ds^2&=r^\frac53\left(-f(r)u_\mu u_\nu dx^\mu dx^\nu+P_{\mu\nu}dx^\mu dx^\nu\right)-2r^\frac16u_\mu dx^\mu dr, \cr
  u^\mu&=\gamma(1,\beta_i),~~\gamma=\frac1{\sqrt{1-\beta_i^2}}.
\end{align}
In the above metric, $u^\mu$ is the four-speed of the relativistic fluid with the normalization $u_\mu u^\mu=-1$. $P_{\mu\nu}$ is the projection tensor of the boundary with $P_{\mu\nu}P^{\nu\rho}=P_\mu^\rho$, which projects any tensor to the plane orthogonal to $u^\mu$. As one can check, Eq.(\ref{eq:boosted 5d bulk}) is the 0th order solution of 5D EOM. The boundary of Eq.(\ref{eq:boosted 5d bulk}) is actually a fluid with constant temperature and velocity, which is of course in global equilibrium.

In order to mimic slightly deviations from local equilibration and the anisotropy of the fluid, we promote the four parameters in Eq.(\ref{eq:boosted 5d bulk}) to be $x^\mu$ dependent: $r_H\to r_H(x),~u_\mu\to u_\mu(x)$, with the requirement that $\big|\frac{\p u}{T}\big|\ll1$, where $T$ is the local temperature of the fluid. Then Eq.(\ref{eq:boosted 5d bulk}) becomes
\begin{align}\label{eq:boosted 5d bulk x dependent}
  ds^2=r^\frac53(-f(r_H(x),r)u_\mu(x) u_\nu(x) dx^\mu dx^\nu+P_{\mu\nu}(x)dx^\mu dx^\nu)-2r^\frac16u_\mu(x) dx^\mu dr,
\end{align}
which is no longer the solution of 5D EOM, but we can make it as the solution again by putting some perturbations in. Using the method of \cite{Bhattacharyya:fluid/gravity}, we should firstly expand the fluid quantities of Eq.(\ref{eq:boosted 5d bulk x dependent}) at some special point, say, $x^\mu=0$ in the local rest frame of the fluid as:
\begin{align}
  u_\mu=-\delta_\mu^0+x^\nu\p_\nu\beta_j\delta^j_\mu,~~~~r_H(x^\mu)=r_H(0)+x^\mu\p_\mu r_H.
\end{align}
$r_H(x^\mu=0)$ is the location of event horizon corresponding to $x^\mu=0$ in the boundary, it relates with the local equilibrium temperature of the fluid at that point. In order to keep the formulations neatly, we will just denote it as $r_H$ in the following calculations but one should always remember that it is a local quantity at $x^\mu=0$. Then we have
\begin{align}
  u_\mu dx^\mu&=-dv+x^\mu\p_\mu\beta_idx^i,~~~~u_\mu u_\nu dx^\mu dx^\nu=dv^2-2x^\mu\p_\mu\beta_idx^idv, \cr
  P_{\mu\nu}dx^\mu dx^\nu&=d\vec x^2-2x^\mu\p_\mu\beta_idx^idv,~~~~f(r_H(x),r)=f(r)-\frac{3r_H^2}{r^3}x^\mu\p_\mu r_H.
\end{align}
Thus Eq.(\ref{eq:boosted 5d bulk x dependent}) becomes
\begin{align}\label{eq:direct 1st od expsn of 0th metric}
  ds^2&=\left(-r^\frac53f+\frac{3r_H^2}{r^\frac43}x^\mu\p_\mu r_H\right)dv^2-\frac{2r_H^3}{r^\frac43}x^\mu\p_\mu\beta_idvdx^i+2r^\frac16dvdr\cr
  &+r^\frac53d\vec x^2-2r^\frac16x^\mu\p_\mu\beta_idx^idr.
\end{align}
The above metric deviates the solution of Einstein equation slightly by the first order boundary derivatives at $x^\mu$, we will see that adding some perturbation terms will make it the solution again, and these perturbations are solved in the next section.

\section{The first order perturbations}

The $SO(3)$ symmetry in Eq.(\ref{eq:5d bulk diagnal metrc}) separates the perturbations into tensor, vector and scalars of $SO(3)$, and we will make use of this advantage to solve these three kinds of perturbations one by one. Generally speaking, all the perturbation ansatz will have the form:
\begin{align}\label{eq:general form of pert}
  P(r) \times
  \begin{cases}
    \p_i\beta_i,~~~~\text{for the scalar part;} \cr
    \p_v\beta_i,~~~~\text{for the vector part;} \cr
    \sigma_{ij},~~~~\text{for the tensor part,}
  \end{cases}
\end{align}
where $P(r)$ is some function of $r$ and can be solved through Einstein equation with the boundary conditions as
\begin{itemize}
  \item $P(r)$ is regular at $r=r_H$;
  \item $\lim_{r\to\infty}\frac{P(r)}{r^n}\to0$.
\end{itemize}
Here $n=0$ or $n=3$ depends on the nature of perturbation terms. We can see that the perturbations will always be of the form of Eq.(\ref{eq:general form of pert}) with the above boundary condition implemented.

\subsection{The tensor part}

We set the tensor part perturbation as
\begin{align}
  ds_{(1)T}^2=r^\frac53\alpha_{ij}(r)dx^idx^j.
\end{align}
The EOM that $\alpha_{ij}$ satisfies is:
\begin{align}
  E_{ij}-\frac13\delta_{ij}\delta^{kl}E_{kl}=T_{ij}-\frac13\delta_{ij}\delta^{kl}T_{kl}.
\end{align}
It turns out that the differential equation for $\alpha_{ij}$ is
\begin{align}\label{eq:1st tensor eom}
  \frac{d}{dr}\left(r^4f\frac{d\alpha_{ij}}{dr}\right)=-5r^\frac32\sigma_{ij},
\end{align}
where $\sigma_{ij}\equiv\p_{(i}\beta_{j)}-\frac13\delta_{ij}\p_k\beta_k$ is the spatial part of shear stress tensor.
The purpose for writing the EOM for tensor part like this is due to the traceless of $\alpha_{ij}$: the trace part of EOM should be removed from the diagonal components. The equation for the 1st order tensor perturbations takes similar form for different models, as can be seen e.g. from \cite{Bhattacharyya:fluid/gravity,Emparan:hydro_black_D3,Erdmenger:hydro_spin_D3}, the reason for this may be due to the universality
 of the shear viscosity in supergravity\footnote{Tensor part perturbation corresponds to the shear viscous term.} \cite{KSS_bound,Buchel:universality_shear}. Since the metric of SS model is also in the class of \cite{Buchel:universality_shear}, so it is natural for Eq. (\ref{eq:1st tensor eom}) to take such a form. We write $\alpha_{ij}$ as $\alpha_{ij}=F(r)\sigma_{ij}$, then $F(r)$ can be solved from
\begin{align}
  F''+\frac{4r^3-r_H^3}{r^4f}F'+\frac5{r^\frac52f}=0,
\end{align}
from which the result can be solved as
\begin{align}
  F(r)&=C_2+\frac1{3\sqrt{r_H}}\left[2\sqrt3\left(\arctan\frac{1-2\sqrt{r/r_H}}{\sqrt3}-\arctan\frac{1+2\sqrt{r/r_H}}{\sqrt3}\right)\right.\cr
   &\left.+\ln\frac{(\sqrt r+\sqrt{r_H})^2(r+\sqrt{rr_H}+r_H)}{(r-\sqrt{rr_H}+r_H)}+C_1\ln\frac{r^3-r_H^3}{r^3}-\ln(\sqrt r-\sqrt{r_H})^2\right].
\end{align}
Regularity at $r=r_H$ requires $C_1=2$ and the normalizability at $r\to\infty$ requires $C_2=2\pi/\sqrt{3r_H}$, thus
\begin{align}
  F(r)&=\frac1{3\sqrt{r_H}}\left[2\sqrt3\left(\arctan\frac{1-2\sqrt{r/r_H}}{\sqrt3}-\arctan\frac{1+2\sqrt{r/r_H}}{\sqrt3}+\pi\right)\right.\cr
  &\left.+\ln\frac{(\sqrt r+\sqrt{r_H})^4(r+\sqrt{rr_H}+r_H)^2(r^2+rr_H+r_H^2)}{r^6}\right].
\end{align}
It is regular at the whole regime of $r>0$ and vanishes to 0 asymptotically.

\subsection{The vector part}

For the vector part, we set the perturbation ansatz as
\begin{align}
  ds_{(1)V}^2=-\frac{2r_H^3}{r^\frac43}w_idx^idv.
\end{align}
The constraint equation for the vector perturbation is
\begin{align}
  g^{rv}(E_{vi}-T_{vi})+g^{rr}(E_{ri}-T_{ri})=0,
\end{align}
which gives
\begin{align}
  \p_ir_H+2r_H\p_v\beta_i=0.
\end{align}
The dynamical equation is
\begin{align}
  E_{ri}-T_{ri}=0.
\end{align}
It turns out that $w_i(r)$ satisfies
\begin{align}
  w_i''-\frac2r w_i'-\frac{5r^\frac12}{2r_H^3}\p_v\beta_i=0,
\end{align}
from which the solution is given as
\begin{align}
  w_i(r)=-\frac{2r^\frac52}{r_H^3}\p_v\beta_i+\frac13r^3C_{1i}+C_{2i}.
\end{align}
It is easy to see that the above general solution is regular at $r_H$. The other boundary condition for the vector part perturbation is
\begin{align}
  \lim_{r\to\infty}\frac{w_i}{r^3}\to 0,
\end{align}
which means $C_{1i}$ must be 0. The appearance of $C_{2i}$ will cause the $(0i)$ components of boundary stress tensor go out of Landau frame. So if one likes to express the boundary stress tensor in Landau frame, $C_{2i}$ should be set to 0. Thus the final result for the vector perturbation of 1st order is
\begin{align}
  ds_{(1)V}^2=4r^\frac76\p_v\beta_idvdx^i.
\end{align}

\subsection{The scalar part}

The scalar part, similar to other works on the effective hydrodynamics of black branes, e.g. \cite{Emparan:hydro_black_D3,Erdmenger:hydro_spin_D3}, is the most complicated part. We set the scalar part perturbation as
\begin{align}
  ds_{(1)S}^2=\frac{k(r)}{r^\frac43}dv^2+r^\frac53h(r)\delta_{ij}dx^idx^j+2r^\frac16j(r)dvdr. \label{eq:scalar perturbt ansatz}
\end{align}
In our case, the gauge condition $\tr[g_{(0)}^{-1}g_{(1)}]=0$ \cite{Bhattacharyya:fluid/gravity,Banerjee:AdS5Q_hydro} cannot be used here for solving the scalar part perturbation, since this will cause inconsistence when solving the EOMs and make the surface stress tensor unrenormalizable. Other gauge condition like $h(r)=1$ \cite{Erdmenger:AdS5Q_hydro} cannot be used either, since the spatial trace part of the metric is non-trivial in the nonconformal case here. Thus we need to keep all the 3 unknowns. However the labor cost by solving all of them gives us a bonus that there will be a bulk viscous term appeared in the surface stress tensor, which doesn't appear in the conformal models with AdS gravity like in Refs.\cite{Bhattacharyya:fluid/gravity,Banerjee:AdS5Q_hydro,Erdmenger:AdS5Q_hydro}. We have two constrain equations for the scalar sector:
\begin{align}
  &g^{rr}(E_{rv}-T_{rv})+g^{rv}(E_{vv}-T_{vv})=0, \\
  &g^{rr}(E_{rr}-T_{rr})+g^{rv}(E_{rv}-T_{rv})=0,
\end{align}
which separately give
\begin{align}
  \p_v r_H=-\frac25 r_H\p_i\beta_i \label{eq:scalar constr 1}
\end{align}
and
\begin{align}
  3(5r^3-2r_H^3)h'-30r^2j-5k'+10r^\frac32\p_i\beta_i=0. \label{eq:scalar constr 2}
\end{align}
We also have a total number of 7 dynamical equations for scalar perturbations, four of them
\begin{align}
  E_{rr}-T_{rr}&=0, \label{eq:scalar EOM rr}\\
  E_{rv}-T_{rv}&=0, \label{eq:scalar EOM rv}\\
  E_{vv}-T_{vv}&=0, \label{eq:scalar EOM vv}\\
  \sum_{i=1}^3 (E_{ii}-T_{ii}) &=0 \label{eq:scalar EOM ii}
\end{align}
come from the Einstein equation Eq.(\ref{eq:5D Ein eq})  and three of them
\begin{align}
  \del^2\phi&-\frac94e^{\frac\phi2-\frac{34}3A+8B}=0, \label{eq:scalar EOM phi} \\
  \del^2A&+\frac{153}{80}e^{\frac\phi2-\frac{34}3A+8B}-\frac{12}5e^{-\frac{16}3A+2B}=0, \label{eq:scalar EOM A}\\
  \del^2B&-\frac9{10}e^{\frac\phi2-\frac{34}3A+8B}+\frac35e^{-\frac{16}3A+2B}=0 \label{eq:scalar EOM B}
\end{align}
are from the 3 scalar field equations in the 5D bulk, namely, Eqs.(\ref{eq:5D phi eq},\ref{eq:5D A eq},\ref{eq:5D B eq}). This looks horrible at a first glance, but fortunately, not all of them give useful message. It turns out that Eq.(\ref{eq:scalar EOM rv}) and Eq.(\ref{eq:scalar EOM vv}) come out of linear compositions of specific constraints with Einstein equation of the scalar sector, so they are not independent equations, and Eqs.(\ref{eq:scalar EOM phi},\ref{eq:scalar EOM A},\ref{eq:scalar EOM B}) give the same differential equation for the 3 unknown scalar perturbations. So we only need to solve Eqs.(\ref{eq:scalar constr 2},\ref{eq:scalar EOM rr},\ref{eq:scalar EOM ii}) and  Eq.(\ref{eq:scalar EOM phi}) to nail Eq.(\ref{eq:scalar perturbt ansatz})  down, among which the last 3 equations are
\begin{align}
  0=~&6rh''+9h'-10j', \label{eq:(rr)}\\
  0=~&12r^4fh''+12(4r^3-r_H^3)h'-6rk''-3k'\cr
       &-6(5r^3-2r_H^3)j'-90r^2j+20r^\frac32\p_i\beta_i, \label{eq:(ii)}\\
  0=~&2r^3fj'+12r^2j+2k'-3r^3fh'-2r^\frac32\p_i\beta_i. \label{eq:(phi)}
\end{align}

We will choose Eqs.(\ref{eq:scalar constr 2},\ref{eq:(rr)},\ref{eq:(phi)}) to solve the 3 unknown scalar perturbations that we set in Eq.(\ref{eq:scalar perturbt ansatz}). From Eq.(\ref{eq:scalar constr 2}) we have
\begin{align}
  6r^2j+k'=\frac35(5r^3-2r_H^3)h'+2r^\frac32\p_i\beta_i,
\end{align}
and after putting it into Eq.(\ref{eq:(phi)}), we get
\begin{align}
  -10r^3fj'=(15r^3+3r_H^3)h'+10r^\frac32\p_i\beta_i.
\end{align}
Then, putting the above equation into Eq.(\ref{eq:(rr)}) one can finally get the equation for $h$:
\begin{align}
  \frac{d}{dr}\left(r^4f\frac{dh}{dr}\right)+\frac53r^\frac32\p_i\beta_i=0.
\end{align}
Without losing of generality, we set $h=F_h(r)\p_i\beta_i$, and $F_h(r)$ satisfies
\begin{align}
  \frac{d}{dr}\left(r^4f\frac{dF_h}{dr}\right)=-\frac53r^\frac32.
\end{align}
If comparing the above equation with Eq.(\ref{eq:1st tensor eom}), one can get $F_h=F/3$ without solving it, thus
\begin{align}
  h=\frac13F(r)\p_i\beta_i.
\end{align}
Inserting $h$ into Eq.(\ref{eq:(rr)}) one has the equation for $j=F_j(r)\p_i\beta_i$ as
\begin{align}
  10F_j'=2rF''+3F'.
\end{align}
This is a 1st order differential equation, and the solution can be obtained by direct integration, the result is
\begin{align}
  F_j(r)=&-\frac25\frac{r^\frac52-r_H^\frac52}{r^3-r_H^3}+C_j+\frac1{30\sqrt{r_H}}\left[2\sqrt3\left(\arctan\frac{1-2\sqrt{r/r_H}}{\sqrt3}
  -\arctan\frac{1+2\sqrt{r/r_H}}{\sqrt3}\right)\right.\cr
  &\left.+\ln\frac{(\sqrt r+\sqrt{r_H})^4(r+\sqrt{rr_H}+r_H)^2(r^2+rr_H+r_H^2)}{r^6}\right].
\end{align}
Since the above expression is already regular at $r=r_H$, the remaining boundary condition for $j$ is
\begin{align}
  \lim_{r\to\infty} F_j \to 0.
\end{align}
Thus $C_j={\sqrt3\pi\over15\sqrt{r_H}}$, so we have finally
\begin{align}
  F_j(r)=-\frac25\frac{r^\frac52-r_H^\frac52}{r^3-r_H^3}+\frac1{10}F.
\end{align}
We set $k=F_k(r)\p_i\beta_i$ likewise and substitute $h$ and $j$ into Eq.(\ref{eq:scalar constr 2}), and we can have
\begin{align}
  F_k(r)=&-\frac{2\sqrt3\pi}{15\sqrt{r_H}}r^3+\frac45r^\frac52+C_k\cr
  &-\frac1{15\sqrt{r_H}}(r^3+2r_H^3)\left[2\sqrt3
  \left(\arctan\frac{1-2\sqrt{r/r_H}}{\sqrt3}-\arctan\frac{1+2\sqrt{r/r_H}}{\sqrt3}\right)\right.\cr
  &\left.+\ln\frac{(\sqrt r+\sqrt{r_H})^4(r+\sqrt{rr_H}+r_H)^2(r^2+rr_H+r_H^2)}{r^6}\right].
\end{align}
The integral constant $C_k$ is fixed by the requirement that the final boundary stress tensor is in the Landau frame, which makes $C_k=-\frac{4\sqrt3\pi}{15}r_H^\frac52$. So we finally have:
\begin{align}
  k=\left(\frac45r^\frac52-\frac15(r^3+2r_H^3)F\right)\p_i\beta_i.
\end{align}
In order to make a consistent check, one may put $h$, $j$ and $k$ into Eq.(\ref{eq:(ii)}), it comes out that the three 1st order scalar perturbations that we have solved out satisfy Eq.(\ref{eq:(ii)}) just right. So the scalar perturbations that we need to make Eq.(\ref{eq:direct 1st od expsn of 0th metric}) as the solution of Einstein equation again turns out to be
\begin{align}
  ds_{(1)S}^2=\left(\frac{F_k}{r^\frac43}dv^2+r^\frac53F_h\delta_{ij}dx^idx^j+2r^\frac16F_jdvdr\right)\p_k\beta_k.
\end{align}

\subsection{Global form of the full metric containing first order perturbations}

Put all the stuff of zeroth and first order together, we get:
\begin{align}
  ds^2&=\left(-r^{5/3}f+\frac{3r_H^2x^\mu\p_\mu r_H}{r^{4/3}}+\frac{F_k\p_i\beta_i}{r^{4/3}}\right)dv^2+\left(4r^{7/6}\p_v\beta_i
  -\frac{2r_H^3x^\mu\p_\mu\beta_i}{r^{4/3}}\right)dx^idv \cr
        &+2r^{1/6}\left(1+F_j\p_i\beta_i\right)dvdr+r^{5/3}\left(\delta_{ij}+\frac13F\delta_{ij}\p_k\beta_k+F\sigma_{ij}\right)dx^idx^j \cr
  &-2r^\frac16x^\mu\p_\mu\beta_idx^idr.
\end{align}
The above is just the full solution of the 1st order at the vicinity of $x^\mu=0$ in some special frame, whose covariant form can be constructed as:
\begin{align}
  ds^2=&-r^\frac53\left(f(r_H(x),r)-\frac{F_k(r_H(x),r)}{r^3}\p_\rho u^\rho\right)u_\mu u_\nu dx^\mu dx^\nu-2r^\frac76(u_\mu a_\nu+u_\nu a_\mu)dx^\mu dx^\nu \cr
        &+r^\frac53F(r_H(x),r)\sigma_{\mu\nu}dx^\mu dx^\nu+r^\frac53\left(1+\frac13F(r_H(x),r)\p_\rho u^\rho\right)P_{\mu\nu}dx^\mu dx^\nu \cr
        &-2r^\frac16\left(1+F_j(r_H(x),r)\p_\rho u^\rho\right)u_\mu dx^\mu dr,
\end{align}
where $\sigma_{\mu\nu}=P_\mu^\rho P_\nu^\sigma\p_{(\rho}u_{\sigma)}-\frac13P_{\mu\nu}\p_\rho u^\rho$ is the 4D covariant shear viscous tensor and $a_\mu=u^\nu\p_\nu u_\mu$ is the 4-acceleration relates with $u_\mu$.

\section{The boundary stress tensor and transport properties}

\subsection{Derivation of boundary stress tensor}

The system of this model is a 5 dimensional Einstein gravity coupled with three scalar fields, its total action can be written as:
\begin{align}
  S=S_{bulk}+S_{GH}+S_{c.t.},
\end{align}
where $S_{bulk}$ is the bulk action (\ref{eq:5D bulk act}), $S_{GH}$ is the corresponding Gibbons-Hawking action:
\begin{align}
  S_{GH}=-\frac1{\kappa_5^2}\int d^4x\sqrt{-h}K,
\end{align}
with $h_{MN}$ is the boundary metric tensor at a hyperplane with constant large $r$. $K$ is the trace of the external curvature. The most crucial part in the total action is the bulk counter term $S_{c.t.}$. Since the bulk metric is not AdS, so the results of the counter term for AdS spacetime \cite{Balasubramanian:stressT_AdS} cannot be directly used here, but fortunately there are also works on the renormalization of nonconformal branes \cite{Kanitscheider:non-confml_holorphy}. Here we adapt the counter term used in \cite{Bigazzi:SS_dyn_flavr}, which they borrow from the much earlier work \cite{D3D7-III} on the renormalization of black D4 brane, in the Einstein frame it has the form of
\begin{align}\label{eq:conter term in Ein frame}
  S_{c.t.}=\frac1{\kappa_{10}^2}\int d^9x\sqrt{-H}\frac52e^{-\frac1{12}\phi}.
\end{align}
Here $H$ is the determinant of the boundary metric of Eq.(\ref{eq:reduct ansatz concrete No.})
\begin{align}
  ds^2=e^{-\frac{10}3A}h_{\mu\nu}dx^\mu dx^\nu+e^{2A+8B}dy^2+e^{2A-2B}d\Omega_4^2.
\end{align}
Note $x^M=\{x^\mu,r\}$. After the dimensional reduction on the above metric, Eq.(\ref{eq:conter term in Ein frame}) becomes
\begin{align}
  S_{c.t.}=\frac1{\kappa_5^2}\int d^4x\sqrt{-h}\frac52e^{-\frac53A-\frac1{12}\phi}.
\end{align}
This counter term contributes to the surface stress tensor as
\begin{align}
  \frac2{\sqrt{-h}}\frac{\delta S_{c.t.}}{\delta h^{\mu\nu}}=\frac1{\kappa_5^2}\left(-\frac52e^{-\frac53A-\frac1{12}\phi}h_{\mu\nu}\right).
\end{align}
Using Eq.(\ref{eq:0th reduc soluti}) and remember that we've set $L=1$, one has the surface stress tensor with contribution from the counter term as
\begin{align}
  T^{surf}_{\mu\nu}=\frac1{\kappa_5^2}\left(K_{\mu\nu}-h_{\mu\nu}K-\frac52r^{-\frac13}h_{\mu\nu}\right).
\end{align}

In the standard technic for 3+1 decomposition of general relativity, $h_{MN}$ is defined as
\begin{align}
  h_{MN}=g_{MN}-n_Mn_N,
\end{align}
where $n_M=N\nabla_M r$ is the unit normal vector for a hyperplane at constant large $r$ in the 5D bulk, of which the metric can be written as
\begin{align}
  ds^2=(N^2+N_MN^M)dr^2+2N_Mdrdx^M+h_{MN}dx^Mdx^N.
\end{align}
$N=(g^{MN}\nabla_M r\nabla_N r)^{-\frac12}$ is called the lapse function and $N^M$ is the shift vector. The index of $n_M$ and $N^M$ goes up and down with $h_{MN}$. The external curvature $K_{MN}$ is related with $h_{MN}$ by
\begin{align}
  K_{MN}=-\frac12\mathcal L_{n} h_{MN}=-\frac12\left(n^P\p_Ph_{MN}+\p_Mn^Ph_{PN}+\p_Nn^Ph_{PM}\right),
\end{align}
in which $\mathcal L_{n}$ is the Lie derivative along the unit normal $n^M$.

\subsection{Transport properties of QGP in D4 holographic QCD model}

The surface stress tensor that we obtain is
\begin{align}\label{eq:boundary stress T}
  T^{surf}_{\mu\nu}=\frac1{2\kappa_5^2}\left(\frac12r_H^3P_{\mu\nu}+\frac52r_H^3u_\mu u_\nu-2r_H^\frac52\sigma_{\mu\nu}
  -\frac4{15}r_H^\frac52\p_\rho u^\rho P_{\mu\nu}\right).
\end{align}
Comparing with the result in relativistic hydrodynamics:
\begin{align}
  T^{hydro}_{\mu\nu}=pP_{\mu\nu}+\varepsilon u_\mu u_\nu-2\eta\sigma_{\mu\nu}-\zeta\p_\rho u^\rho P_{\mu\nu},
\end{align}
where $p$, $\varepsilon$, $\eta$ and $\zeta$ are the momentum density, the energy density, shear viscosity and bulk viscosity, respectively. We can get the respective hydrodynamical quantities for our system as
\begin{align}\label{eq:hydrodynamical_quantities_SS}
  p=\frac1{2\kappa_5^2}\frac12r_H^3,~~~~\varepsilon=\frac1{2\kappa_5^2}\frac52r_H^3,~~~~\eta=\frac1{2\kappa_5^2}r_H^\frac52,
  ~~~~\zeta=\frac1{2\kappa_5^2}\frac4{15}r_H^\frac52.
\end{align}
From Eq.(\ref{eq:5d bulk diagnal metrc}) we can get the temperature for the 5D spacetime as
\begin{align}
  T=\frac{3r_H^\frac12}{4\pi}.
\end{align}
As one can easily seen from the above two expressions, both the thermodynamic and the transport coefficients are only depend on temperature. This is due to the setup of this model. One can also get the entropy density as
\begin{align}
  s=\frac{\varepsilon+p}T=\frac1{2\kappa_5^2}4\pi r_H^\frac52.
\end{align}
So the ratios of shear and bulk viscosity to entropy density are
\begin{align}
  \frac\eta s=\frac1{4\pi},~~~~\frac\zeta s=\frac1{15\pi}.
\end{align}
Here we meet the renowned $1/4\pi$ again and this suggests both the bulk of D4-D6 model and SS model belong to the class in \cite{Buchel:universality_shear}, as \cite{Benincasa:hydro_SS} has pointed out. The bulk to shear ratio is
\begin{align}\label{bulk to shear SS}
  \frac\zeta\eta=\frac4{15},
\end{align}
which is also the same as in \cite{Benincasa:hydro_SS} and \cite{Eling:novel_formla_bulk_viscs}. It is interesting to compare our result Eq.(\ref{bulk to shear SS}) with the results of Refs.\cite{Mas:Dp_hydro_GreenKubo,Skenderis:Dp-brane_hydro} of which for near-horizon non-extremal D4-brane is $1/10$. This is understandable since the case we considered here is the compactified near-horizon, non-extremal D4-brane in which the relativistic fluid resides only on 1+3 dimensions out of the 1+4 dimensional D4-brane's world-volume. The spacetime here comes from dimensional reduction on $S^1\times S^4$. But in Refs.\cite{Mas:Dp_hydro_GreenKubo,Skenderis:Dp-brane_hydro}, for D4-brane, the submanifold that is reduced is the $S^4$ and the relevant hydrodynamics is 1+4 dimensional. Another consistency with \cite{Benincasa:hydro_SS} is the sound speed that can be obtained via thermodynamic quantities:
\begin{align}\label{eq:sound speed}
  c_s^2=\frac{\p p}{\p\varepsilon}=\frac15.
\end{align}

As a self-consistent check, we calculate the dispersion relations by using the constituent relation Eq.(\ref{eq:boundary stress T}) as in \cite{Bhattacharyya:fluid/gravity}. If considering the temperature $r_H(x)$ and 3-velocity $\beta_i(x)$ has fluctuations as
\begin{align}
  r_H(x)=r_H+\delta r_H e^{-i\omega v+i\vec k\cdot\vec x},~~~~\beta_i(x)=\delta\beta_i e^{-i\omega v+i\vec k\cdot\vec x},
\end{align}
one can get the relations of the fluctuations by putting the above equations into the EOM of boundary fluid, i.e., the conservation equation for $T^{surf}_{\mu\nu}$:
\begin{align}
  \p^\mu T^{surf}_{\mu\nu}=0.
\end{align}
Treating $\delta r_H$ and $\delta\beta_i$ as first order quantities, one can get the linear equation for the fluctuations
\begin{align}
  &\frac52\omega\delta r_H-r_Hk_i\delta\beta_i=0, \\
  &\frac{3i}2k_i\delta r_H+(r_H^{1/2}{\vec k}^2-3ir_H\omega)\delta\beta_i+\frac35r_H^{1/2}k_ik_j\delta\beta_j=0.
\end{align}
In order to make the above equations have non-trivial solution, the determinant of coefficients should be 0, which gives
\begin{align}
  \omega&=-\frac i{3r_H^{1/2}}\vec k^2,~~~~\text{shear mode} \cr
  \omega&=\pm\frac1{\sqrt5}|\vec k|-i\frac4{15r_H^{1/2}}\vec k^2+\mathcal O(|\vec k|^3).~~~~\text{sound mode}
\end{align}
Comparing with the following results in hydrodynamics
\begin{align}
  \omega&=-i\frac\eta{\varepsilon+p}\vec k^2,~~~~\text{shear mode} \cr
  \omega&=c_s |\vec k|-i\frac{\zeta+\frac43\eta}{2(\varepsilon+p)}\vec k^2.~~~~\text{sound mode}
\end{align}
One can read the following relations
\begin{align}
  \frac\eta{\varepsilon+p}=\frac1{3r_H^{1/2}},~~~~c_s^2=\frac15,~~~~\frac{\zeta+\frac43\eta}{2(\varepsilon+p)}=\frac4{15r_H^{1/2}}.
\end{align}
Comparing with the results in Eq.(\ref{eq:hydrodynamical_quantities_SS}), we can find perfect consistency.

\section{Discussions and outlooks}

We develop the BDE formalism of fluid/gravity correspondence in compactified black D4-brane background and investigate the transport properties of its gauge-side dual gluonic matter. Compactified D4 branes are background of the D4-D6 and the SS models, which are the two nonconformal top-down holographic QCD models. The SS model is a holographic model whose dual field theory lives on the world-volume of the flavor D8 branes, it is convenient to extract hadronic properties such as the meson and baryon spectrums from the SS model, since the DBI action of D8 branes describes some meson effective theory like the $\chi$-PT. However, in the SS model, people focus more on the flavor sector, and may ignore the bulk sector. Our current work focuses on the bulk sector, i.e., the compactified black D4 brane background whose asymptotic region is not an AdS spacetime. Therefore, we choose the compactified black D4 brane background to describe nonconformal gluonic matter.

The strategy is to use the dimensional reduction technique on the compact structure of SS model background in Einstein frame, and one can get a 5D effective Einstein gravity minimally coupled with 3 scalars with exponential potentials. Following the standard BDE formalism of fluid/gravity correspondence, we derive the constituent relation and read the thermodynamical and hydrodynamical quantities such as the energy and momentum density and the shear and bulk viscosities. It is found that the ratio of bulk to shear viscosity and sound speed from our results are consistent with the previous studies on the transport properties of SS model \cite{Benincasa:hydro_SS,Eling:novel_formla_bulk_viscs}, which shows the validity of the BDE formalism of fluid/gravity correspondence in nonconformal background. The calculation of second order transport coefficients are technically direct based on this work. What's more, this work offers us an nonconformal prototype in fluid/gravity duality that is in parallel with Bharttacharyya et al.'s $AdS_5$ construction, which provides us the opportunity to study nonconformal systems by the 5D metric (\ref{eq:5d bulk diagnal metrc}) given in this work.


As further applications, the most straightforward project is to calculate the second order transport coefficients of nonconformal gluonic matter by using the BDE formalism. The second order transport coefficients have been calculated in conformal systems with \cite{Banerjee:AdS5Q_hydro,Erdmenger:AdS5Q_hydro} or without \cite{Bhattacharyya:fluid/gravity,AdS5_dilaton_hydro} chemical potentials using the BDE formalism of fluid/gravity duality, and also been calculated via Green-Kubo formulae in both conformal \cite{Baier:AdS/CFT_hydro_2order} and nonconformal \cite{Finazzo:transpt_coeficit_non-conf_QGP} systems. For the nonconformal Dp-brane backgrounds, the form of second order viscous tensor has been predicted in \cite{Skenderis:Dp-brane_hydro}.

One may also study the effective fluid on a cavity with finite $r=R$ using the compactified black D4 brane solution (not under the near horizon limit) like \cite{Emparan:hydro_black_D3,Erdmenger:hydro_spin_D3}. The most different point of the effective recipe from the present work is the dilaton, $A$ and  $B$ should be boundary dependent since they both relate with $H_4$, and thus $r_{Q4}$. This is like the case in \cite{Erdmenger:hydro_spin_D3} where
scalars also relate with the harmonic functions, but unlike that in \cite{Emparan:hydro_black_D3}, in which scalar is only $r$-dependent and the cut-off surface can be chosen as iso-dilatonic. Thus we should let those 3 scalars all have 1st order perturbations just like in \cite{Erdmenger:hydro_spin_D3}, which may make us solve 6 equations in all for the scalar part perturbation.

Another interesting attempt in the future is to investigate nonconformal fluid with an axial chemical potential $\mu_5$ via the SS model with smeared D0 charge on the D4 brane world-volume \cite{Wuchao:D0D4_SS}, this model can extract the axial chemical potential and the axial charge diffusion constant besides the hydrodynamical quantities in the present work. What's more, all the hydrodynamical quantities should be both temperature and the axial chemical potential dependent. We can also use the method developed in this work to investigate the newly found anomalous effects \cite{LiaoJinfeng:anomalous_effect_HIC} (such as the Chiral Magnetic Effect, Chiral Separation Effect and so on) analytically.

\vskip 1 cm
{\bf Acknowledgement.---}
We shall thank Yu Lu for his great help on the calculation code, and we'd also like to thank Soeren Schlichting, Wilke van der Schee, Jiaju Zhang and Zhenhua Zhou for their helpful discussions on the technical and physical problems about this work. This work is supported by the NSFC under
Grant No. 11275213, and 11261130311(CRC 110 by DFG and NSFC), CAS key project KJCX2-EW-N01, and Youth
Innovation Promotion Association of CAS.

\appendix
\section{Dimensional reduction from 10D to 5D}
\label{Apendix}

We will do dimensional reduction on the following 10D diagonal background:
\begin{align}
  ds^2=e^{2\alpha_1A}g_{MN}dx^Mdx^N+e^{2\alpha_2A}\left( e^{2\beta_1B}dy^2
         +e^{2\beta_2B}\gamma_{ab}d\theta^ad\theta^b \right).
\end{align}
The non-trivial Christoffel symbol of this metric are (the ones with ``tilde" are 10 dimensional components):
\begin{align}
  \tilde\Gamma^M_{NP}&=\Gamma^M_{NP}+\alpha_1( \delta^M_N\p_P A+\delta^M_P\p_N A-g_{NP}\nabla^M A ); \cr
  \tilde\Gamma^M_{yy}&=-(\alpha_2\nabla^M A+\beta_1\del^M B)e^{(-2\alpha_1+2\alpha_2)A+2\beta_1B}; \cr
  \tilde\Gamma^y_{My}&=\alpha_2\p_M A+\beta_1\p_M B; \cr
  \tilde\Gamma^M_{ab}&=-(\alpha_2\del^M A+\beta_2\del^M B)e^{(-2\alpha_1+2\alpha_2)A+2\beta_2B}\gamma_{ab}; \cr
  \tilde\Gamma^a_{Mb}&=(\alpha_2\p_M A+\beta_2\p_M B)\delta^a_b; \cr
  \tilde\Gamma^a_{bc}&=\Gamma^a_{bc}.
\end{align}
From the above results, we also have:
\begin{align}
  \tilde\Gamma^N_{MN}&=\Gamma^N_{MN}+5\alpha_1\p_M A, \cr
  \tilde\Gamma^P_{MP}+\tilde\Gamma^y_{My}+\tilde\Gamma^a_{Ma}&=\Gamma^P_{MP}+(5\alpha_1+5\alpha_2)\p_M A
  +(\beta_1+4\beta_2)\p_M B,
\end{align}
which can make our computation more convenient. The components of Ricci tensors are:
\begin{align}
  \tilde R_{MN}=&R_{MN}-(3\alpha_1+5\alpha_2)\del_M\del_N A-(\beta_1+4\beta_2)\del_M\del_N B-\alpha_1
                                              g_{MN}\del_P\del^P A \cr
               &+(3\alpha_1^2+10\alpha_1\alpha_2-5\alpha_2^2)\p_M A\p_N A-(3\alpha_1^2+5\alpha_1\alpha_2)g_{MN}(\p A)^2 \cr
               &+(\alpha_1\beta_1-\alpha_2\beta_1+4\alpha_1\beta_2-4\alpha_2\beta_2)(\p_M A\p_N B+\p_N A\p_M B) \cr
               &-(\alpha_1\beta_1+4\alpha_1\beta_2)g_{MN}\p_PA\p^PB-(\beta_1^2+4\beta_2^2)\p_MB\p_NB; \\
  \tilde R_{yy}=&-\left[ \alpha_2\del^2A+\beta_1\del^2B+(3\alpha_1\alpha_2+5\alpha_2^2)(\p A)^2
                                    +(3\alpha_1\beta_1+6\alpha_2\beta_1+4\alpha_2\beta_2)\p A\cdot\p B \right. \cr
                          &\left.+(\beta_1^2+4\beta_1\beta_2)(\p B)^2 \right]e^{(-2\alpha_1+2\alpha_2)A+2\beta_1B}; \\
  \tilde R_{ab}=&-\left[ \alpha_2\del^2 A+\beta_2\del^2 B+(3\alpha_1\alpha_2+5\alpha_2^2)(\p A)^2
                                        +(3\alpha_1\beta_2+\alpha_2\beta_1+9\alpha_2\beta_2)\p A\cdot\p B \right. \cr
                         &\left. +(\beta_1\beta_2+4\beta_2^2)(\p B)^2 \right]e^{(-2\alpha_1+2\alpha_2)A+2\beta_2B}\gamma_{ab}
                                     +3\gamma_{ab}.
\end{align}
Again, the components with ``tilde" are 10 dimensional ones. Then we have the Ricci scalar:
\begin{align}
  R^{(10)}=&\left[ R-(8\alpha_1+10\alpha_2)\del^2A-(2\beta_1+8\beta_2)\del^2B-(12\alpha_1^2+30\alpha_1\alpha_2+30\alpha_2^2)(\p A)^2\right.\cr
              &\left. -6(\alpha_1+2\alpha_2)(\beta_1+4\beta_2)\p A\cdot\p B-(2\beta_1^2+8\beta_1\beta_2+20\beta_2^2)(\p B)^2 \right]e^{-2\alpha_1A}\cr
              &+12e^{-2\alpha_2A-2\beta_2B}.
  \label{reduction of Ricci Scalar}
\end{align}


\begin{thebibliography}{99}

\bibitem{RHIC-EXP}
I.~Arsene {\it et al.} [BRAHMS Collaboration], ``Quark gluon plasma and color glass condensate at RHIC? The perspective from the BRAHMS experiment,'' Nucl.\ Phys.\  A {\bf 757} (2005) 1, [arXiv:nucl-ex/0410020];
K.~Adcox {\it et al.} [PHENIX Collaboration], ``Formation of dense partonic matter in relativistic nucleus nucleus collisions at RHIC: Experimental evaluation by the PHENIX collaboration,'' Nucl.\ Phys.\  A {\bf 757} (2005) 184, [arXiv:nucl-ex/0410003];
B.~B.~Back {\it et al.}, ``The PHOBOS perspective on discoveries at RHIC,'' Nucl.\ Phys.\  A {\bf 757} (2005) 28, [arXiv:nucl-ex/0410022];
J.~Adams {\it et al.} [STAR Collaboration], ``Experimental and theoretical challenges in the search for the quarku gluon plasma: The STAR collaboration's critical assessment of the evidence from RHIC collisions,'' Nucl.\ Phys.\  A {\bf 757} (2005) 102, [arXiv:nucl-ex/0501009].

\bibitem{RHIC-THEO}
M.~Gyulassy and L.~McLerran, ``New forms of QCD matter discovered at RHIC,'' Nucl.\ Phys.\  A {\bf 750}, 30 (2005), [arXiv:nucl-th/0405013].

\bibitem{KSS_bound}
P. Kovtun, D. T. Son and A. O. Starinets, ``Holography and hydrodynamics: diffusion on stretched horizons'', JHEP \textbf{0310} (2003) 064, [arXiv:hep-th/0309213].

\bibitem{Kovtun:2004de}
P.~Kovtun, D.~T.~Son and A.~O.~Starinets, ``Viscosity in strongly interacting quantum field theories from black hole physics'', Phys.\ Rev.\ Lett.\  {\bf 94}, 111601 (2005), [arXiv:hep-th/0405231].

\bibitem{Song:2008hj}
H.~Song and U.~W.~Heinz, ``Extracting the QGP viscosity from RHIC data---A Status report from viscous hydrodynamics,''
J.\ Phys.\ G {\bf 36} (2009) 064033, [arXiv:0812.4274 [nucl-th]].

\bibitem{Hydro}
D.~Teaney, J.~Lauret and E.~V.~Shuryak, ``Flow at the SPS and RHIC as a quark gluon plasma signature,'' Phys.\ Rev.\ Lett.\  {\bf 86} 4783 (2001), [arXiv:nucl-th/0011058];
P.~Huovinen, P.~F.~Kolb, U.~W.~Heinz, P.~V.~Ruuskanen and S.~A.~Voloshin, ``Radial and elliptic flow at RHIC: Further predictions,'' Phys.\ Lett.\  B {\bf 503} (2001) 58, [arXiv:hep-ph/0101136];
T.~Hirano, U.~W.~Heinz, D.~Kharzeev, R.~Lacey and Y.~Nara, ``Hadronic dissipative effects on elliptic flow in ultrarelativistic heavy-ion collisions,'' Phys.\ Lett.\  B {\bf 636} (2006) 299, [arXiv:nucl-th/0511046];
P.~Romatschke and U.~Romatschke, ``Viscosity Information from Relativistic Nuclear Collisions: How Perfect is the Fluid Observed at RHIC?'', Phys.\ Rev.\ Lett.\  {\bf 99} (2007) 172301, [arXiv:0706.1522 [nucl-th]];
H.~Song and U.~W.~Heinz, ``Suppression of elliptic flow in a minimally viscous quark-gluon plasma,'' Phys.\ Lett.\  B {\bf 658} (2008) 279,  [arXiv:0709.0742 [nucl-th]];
H.~Song and U.~W.~Heinz, ``Causal viscous hydrodynamics in 2+1 dimensions for relativistic heavy-ion collisions,'' Phys.\ Rev.\  C {\bf 77} (2008) 064901, [arXiv:0712.3715 [nucl-th]].

\bibitem{Hydro-Teaney}
D.~Teaney, ``Effect of shear viscosity on spectra, elliptic flow, and Hanbury Brown-Twiss radii,'' Phys.\ Rev.\  C {\bf 68}, 034913 (2003), [arXiv:nucl-th/0301099].

\bibitem{Maldacena:AdS/CFT}
J. M. Maldacena, ``\emph{The Large N Limit of Superconformal Field Theories and Supergravity}", Adv. Theor. Math. Phys. {\bf 2} (1998) 231, [arXiv:hep-th/9711200].

\bibitem{GKP:AdS/CFT}
S. S. Gubser, I. R. Klebanov and A. M. Polyakov, ``\emph{Gauge Theory Correlators from Non-critical String Theory}", Phys. Lett. {\bf B 428} (1998) 105, [arXiv:hep-th/9802109].

\bibitem{Witten:AdS/CFT}
E. Witten, ``\emph{Anti-de Sitter Space and Holography}", Adv. Theor. Math. Phys. {\bf 2} (1998) 253, [arXiv:hep-th/9802150].

\bibitem{near_horizon_struct1}
G. Gibbons, ``\emph{Antigravitating Black Hole Solitons with Scalar Hair in $\mathcal N=4$ Supergravity}", Nucl. Phys. {\bf B 207} (1982) 337.

\bibitem{near_horizon_struct2}
R. Kallosh and A. Peet, ``\emph{Dilaton Black Holes Near the Horizon}", Phys. Rev. {\bf D 46} (1992) 5223, [arXiv:hep-th/9209116].

\bibitem{near_horizon_struct3}
G. Gibbons and P. Townsend, ``\emph{Vacuum Interpolation In Supergravity via Super $p$-Branes}", Phys. Rev. Lett. {\bf 71} (1993) 3754, [arXiv:hep-th/9307049].

\bibitem{near_horizon_struct4}
M. J. Duff, G. W. Gibbons, and P. K. Townsend, ``\emph{Macroscopic Superstrings As Interpolating Solitons}", Phys. Lett. {\bf B 332} (1994) 321, [arXiv:hep-th/9405124].

\bibitem{near_horizon_struct5}
G. W. Gibbons, G. T. Horowitz, and P. K. Townsend, ``\emph{Higher Dimensional Resolution Of Dilatonic Black Hole Singularities}", Class. Quant. Grav. {\bf 12} (1995) 297, [arXiv:hep-th/9410073].

\bibitem{near_horizon_struct6}
S. Ferrara, G. W. Gibbons, and R. Kallosh, ``\emph{Black Holes And Critical Points In Moduli Space}", Nucl. Phys. {\bf B 500} (1997) 75, [arXiv:hep-th/9702103].

\bibitem{near_horizon_struct7}
A. Chamseddine, S. Ferrara, G. W. Gibbons, and R. Kallosh,``\emph{Enhancement Of Supersymmetry Near 5D Black Hole Horizon}", Phys. Rev. {\bf D 55} (1997) 3647, [arXiv:hep-th/9610155].

\bibitem{brane_absorption1}
I. R. Klebanov, ``\emph{World Volume Approach To Absorption By Nondilatonic Branes}", Nucl. Phys. {\bf B 496} (1997) 231, [arXiv:hep-th/9702076].

\bibitem{brane_absorption2}
S. S. Gubser, I. R. Klebanov, A. A. Tseytlin, ``\emph{String Theory And Classical Absorption By Three-branes}", Nucl. Phys. {\bf B 499} (1997) 217, [arXiv:hep-th/9703040].

\bibitem{brane_absorption3}
S. S. Gubser and I. R. Klebanov, ``\emph{Absorption By Branes And Schwinger Terms In The World Volume Theory}", Phys. Lett. {\bf B 413} (1997) 41, [arXiv:hep-th/9708005].

\bibitem{brane_absorption4}
J. Maldacena and A. Strominger, ``\emph{Semiclassical Decay Of Near Extremal Fivebranes}'', JHEP {\bf 9712} (1997) 008, [arXiv:hep-th/9710014].

\bibitem{Itzhaki4:gauge/gravity}
N. Itzhaki, J. M. Maldacena, J. Sonnenschein and S. Yankielowicz, ``\emph{Supergravity and the Large N Limit of Theories with Sixteen Supercharges}", Phys. Rev. {\bf D 58} (1998) 046004, [arXiv:hep-th/9802042].

\bibitem{Earliest_Glueball1}
D. J. Gross and H. Ooguri, ``\emph{Aspects of Large N Gauge Theory Dynamics as Seen by String Theory}", Phys. Rev. {\bf D 58} (1998) 106002, [arXiv:hep-th/9805129].

\bibitem{Earliest_Glueball2}
C. Cs\'aki, H. Ooguri, Y. Oz and J. Terning, ``\emph{Glueball Mass Spectrum from Supergravity}", JHEP \textbf{9901} (1999) 017, [arXiv:hep-th/9806021].

\bibitem{Earliest_Glueball3}
R. de Mello Koch, A. Jevicki, M. Mihailescu and J. P. Nunes, ``\emph{Evaluation of Glueball Masses from Supergravity}'', Phys. Rev. \textbf{D 58} (1998) 105009, [arXiv:hep-th/9806125].

\bibitem{D3D7-I}
M. Kruczenski, D. Mateos, R. C. Myers and D. J. Winters, ``\emph{Meson Spectroscopy in AdS/CFT with Flavor}", JHEP \textbf{0307} (2003) 049, [arXiv:hep-th/0304032].

\bibitem{D3D7-II}
D. Mateos, R. C. Myers and R. M. Thomson, ``\emph{Holographic Phase Transitions with Fundamental Matter}", Phys. Rev. Lett. \textbf{97} (2006) 091601, [arXiv:hep-th/0605046].

\bibitem{D3D7-III}
D. Mateos, R. C. Myers and R. M. Thomson, ``\emph{Thermodynamics of the Brane}", JHEP \textbf{0705} (2007) 067, [arXiv:hep-th/0701132].

\bibitem{D4D6}
M. Kruczenski, D. Mateos, R. C. Myers and D. J. Winters, ``\emph{Towards a Holographic Dual of Large $N_c$ QCD}", JHEP \textbf{0405} (2004) 041, [arXiv:hep-th/0311270].

\bibitem{phase_transit_SS}
O. Aharony, J. Sonnenschein and S. Yankielowicz, ``A Holographic Model of Deconfinement and Chiral Symmetry Restoration", Annals Phys. \textbf{322} (2007) 1420, [arXiv:hep-th/0604161].

\bibitem{D4D8}
T. Sakai and S. Sugimoto, ``\emph{Low Energy Hadron Physics in Holographic QCD}", Prog. Theor. Phys. \textbf{113} (2005) 843,
[arXiv:hep-th/0412141].

\bibitem{Adding_Flavor1}
A. Karch and E. Katz, ``\emph{Adding Flavor to AdS/CFT}", JHEP \textbf{0206} (2002) 043, [arXiv:hep-th/0205236].

\bibitem{Adding_Flavor2}
A. Karch and Randall, ``\emph{Open and Closed String Interpretation of SUSY CFT¡¯s on Branes with Boundaries}", JHEP \textbf{0106} (2001) 063, [arXiv:hep-th/0105132].

\bibitem{Adding_Flavor3}
O. Aharony, A. Fayyazuddin and J. M. Maldacena, ``\emph{The Large N Limit of N = 2, 1 Field Theories from Three-branes in F-theory}", JHEP \textbf{9807} (1998) 013, [arXiv:hep-th/9806159].

\bibitem{Policastro:eta_N=4_SYM_plasma}
G. Policastro, D. T. Son and A. O. Starinets, ``Shear viscosity of strongly coupled $\mathcal N=4$ supersymmetric Yang-Mills plasma", Phys. Rev. Lett. \textbf{87} (2001) 081601, [arXiv:hep-th/0104066].

\bibitem{Son:real_time_AdS/CFT}
D. T. Son and A. O. Starinets, ``Minkowski-space correlators in AdS/CFT correspondence: recipe and applications", JHEP \textbf{0209} (2002) 042, [arXiv:hep-th/0205051].

\bibitem{Kubo_formulae}
R. Kubo,``Statistical Mechanical Theory of Irreversible Proceses'', Journal of the Physical Society of Japan Vol. \textbf{12} (1957) 570.

\bibitem{Hosaya:Kubo_form_particle_phys}
A. Hosoya, M. Sakagami and M. Takao, ``Nonequilibrium Thermodynamics in Field Theory: Transport Coefficients'', Ann. Phys. \textbf{154} (1984) 229.

\bibitem{Policastro:AdS/CFT_hydro1}
G. Policastro, D. T. Son and A. O. Starinets, ``From AdS/CFT correspondence to hydrodynamics", JHEP \textbf{0209} (2002) 043, [arXiv:hep-th/0205052].

\bibitem{Policastro:AdS/CFT_hydro2}
G. Policastro, D. T. Son and A. O. Starinets, ``From AdS/CFT correspondence to hydrodynamics. II. Sound waves", JHEP \textbf{0212} (2002) 054, [arXiv:hep-th/0210220].

\bibitem{Baier:AdS/CFT_hydro_2order}
R. Baier, P. Romatschke, D. T. Son, A. O. Starinets and M. A. Stephanov, ``Relativistic viscous hydrodynamics, conformal invariance and holography", JHEP \textbf{0804} (2008) 100, [arXiv:0712.2451 [hep-th]].


\bibitem{Son;1order_hydro_review}
D. T. Son and A. O. Starinets, ``Viscosity, black holes and quantum field theory", Ann. Rev. Nucl. Part. Sci. \textbf{57} (2007) 95, [arXiv:0704.0240 [hep-th]].

\bibitem{Bhattacharyya:fluid/gravity}
S. Bhattacharyya, V. E. Hubeny, S. Minwalla and M. Rangamani, ``Nonlinear fluid dynamics from gravity", JHEP \textbf{0802} (2008) 045, [arXiv:0712.2456 [hep-th]].

\bibitem{Bhattacharyya:Js}
S. Bhattacharyya, V. E. Hubeny, R. Loganayagam, G. Mandal, S. Minwalla, T. Morita and M. Rangamani, ``Local Fluid Dynamical Entropy from Gravity",
JHEP \textbf{0806} (2008) 055, [arXiv:0803.2526 [hep-th]].

\bibitem{AdS5_dilaton_hydro}
S. Bhattacharyya, R. Loganayagam, S. Minwalla, S. Nampuri, S. P. Trivedi and S. R. Wadia, ``Forced fluid dynamics from gravity", JHEP \textbf{0902} (2009) 018, [arXiv:0806.0006 [hep-th]].

\bibitem{Banerjee:AdS5Q_hydro}
N. Banerjee, J. Bhattacharya, S. Bhattacharyya, S. Dutta, R. Loganayagam and P. Surowka, ``Hydrodynamics from charged black holes", JHEP \textbf{1101} (2011) 094, [arXiv:0809.2596 [hep-th]].

\bibitem{Erdmenger:AdS5Q_hydro}
J. Erdmenger, M. Haack, M. Kaminski and A. Yarom, ``Fluid dynamics of R-charged black holes'', JHEP \textbf{0901} (2009) 055, [arXiv:0809.2488 [hep-th]].

\bibitem{blackfold1}
R. Emparan, T. Harmark, V. Niarchos and N. A. Obers, ``World-volume effective theory for higher-dimensional black holes'', Phys. Rev. Lett. \textbf{102}
(2009) 191301, [arXiv:0902.0427 [hep-th]].

\bibitem{blackfold2}
R. Emparan, T. Harmark, V. Niarchos and N. A. Obers, ``Essentials of blackfold dynamics'', JHEP \textbf{1005} (2010) 042, [arXiv:0910.1610 [hep-th]].

\bibitem{blackfold3}
J. Camps, R. Emparan and N. Haddad, ``Black Brane Viscosity and the Gregory Laflamme Instability", JHEP \textbf{1005} (2010) 042, [arXiv:1003.3636 [hep-th]].

\bibitem{blackfold4}
J. Camps and R. Emparan, ``Derivation of the blackfold effective theory'', JHEP \textbf{1203} (2012) 038, [arXiv:1201.3506 [hep-th]].

\bibitem{Emparan:blackfold_hydro}
R. Emparan and M. Mart\'inez, ``Black Branes in a Box: Hydrodynamics, Stability and Criticality", JHEP \textbf{1207} (2012) 120, [arXiv:1205.5646 [hep-th]].

\bibitem{Emparan:hydro_black_D3}
R. Emparan, V. E. Hubeny and M. Rangamani, ``Effective hydrodynamics of black D3-branes", JHEP \textbf{1306} (2013) 035, [arXiv:1303.3563 [hep-th]].

\bibitem{Erdmenger:hydro_spin_D3}
J. Erdmenger, M. Rangamani, S. Steinfurt and H. Zeller, ``Hydrodynamic regimes of spinning black D3 branes", JHEP \textbf{1502} (2015) 026, [arXiv:1412.0020 [hep-th]].

\bibitem{LiuHong:hot_QCD}
Casalderrey-Solana, Hong Liu, D. Mateos, K. Rajagopal and U. A. Wiedemann, ``Gauge String Duality, Hot QCD and Heavy Ion Collions", [arXiv:1101.0618 [hep-th]].

\bibitem{Iancu:hearvy_ion}
E. Iancu, ``QCD in Heavy Ion Collisions", 2011 European School of High-Energy Physics, Cheile Gradistei, Romania, 07-20 Sep. 2011 pp.197-266, [arXiv:1205.0579 [hep-th]].

\bibitem{LAT-etas}
A.~Nakamura and S.~Sakai, ``Transport coefficients of gluon plasma'', Phys.\ Rev.\ Lett.\  {\bf 94} (2005) 072305, [arXiv:hep-lat/0406009].

\bibitem{LAT-xis-KT}
D.~Kharzeev and K.~Tuchin, ``Bulk viscosity of QCD matter near the critical temperature,'' JHEP \textbf{0809} (2008) 093, [arXiv:0705.4280 [hep-ph]];
F.~Karsch, D.~Kharzeev and K.~Tuchin, ``Universal properties of bulk viscosity near the QCD phase transition,'' Phys.\ Lett.\  B {\bf 663} (2008) 217, [arXiv:0711.0914 [hep-ph]].

\bibitem{LAT-xis-Meyer}
H.~B.~Meyer, ``A calculation of the bulk viscosity in SU(3) gluodynamics,'' Phys.\ Rev.\ Lett.\  {\bf 100} (2008) 162001, [arXiv:0710.3717 [hep-lat]].

\bibitem{correlation-Karsch}
  K.~Huebner, F.~Karsch and C.~Pica, ``Correlation functions of the energy-momentum tensor in SU(2) gauge theory at finite temperature,'' Phys. Rev. \textbf{D 78} (2008) 094501, [arXiv:0808.1127 [hep-lat].

\bibitem{bottomup-gubser}
  S.~S.~Gubser and A.~Nellore, ``Mimicking the QCD equation of state with a dual black hole,'' Phys.\ Rev.\ D {\bf 78} (2008) 086007, [arXiv:0804.0434 [hep-th]].  

\bibitem{bottomup-kampfer}
R.~Yaresko and B.~Kampfer, ``Bulk viscosity of the gluon plasma in a holographic approach,'' Acta Phys.\ Polon.\ Supp.\ {\bf 7} (2014) no.1 137, [arXiv:1403.3581 [hep-ph]].

\bibitem{bottomup-noronha}
J.~Noronha-Hostler, J.~Noronha and F.~Grassi, ``Bulk viscosity-driven suppression of shear viscosity effects on the flow harmonics at energies available at the BNL Relativistic Heavy Ion Collider,'' Phys.\ Rev.\ C {\bf 90} (2014) 3, 034907, [arXiv:1406.3333 [nucl-th]].

\bibitem{bottomup-huangli}
D.~Li, S.~He and M.~Huang, ``Temperature dependent transport coefficients in a dynamical holographic QCD model,'' JHEP {\bf 1506} (2015) 046,  [arXiv:1411.5332 [hep-ph]].

\bibitem{David:D1-brane_hydro}
J. R. David, M. Mahato and S. R. Wadia, ``Hydrodynamics from the D1-brane'', [arXiv:0901.2013].

\bibitem{Mas:Dp_hydro_GreenKubo}
J. Mas and J. Tarr\'io, ``Hydrodynamics from the Dp-brane", JHEP \textbf{0705} (2007) 036, [arXiv:hep-th/0703093].

\bibitem{Skenderis:Dp-brane_hydro}
I. Kanitscheider and K. Skenderis, ``Universal hydrodynamics of nonconformal branes", JHEP \textbf{0904} (2009) 062, [arXiv:0901.1487].

\bibitem{Gupta:BDE_FeffermanGraham}
R. K. Gupta and A. Mukhopadhyay, ``On the universal hydrodynamics of strongly coupled CFTs with gravity duals", JHEP \textbf{0903} (2009) 067, [arXiv:0810.4851].

\bibitem{Benincasa:hydro_SS}
P. Benincasa and A. Buchel, ``Hydrodynamics of Sakai-Sugimoto model in the quenched approximation", Phys. Lett. \textbf{B 640} (2006) 108, [arXiv:hep-th/0605076].

\bibitem{Eling:novel_formla_bulk_viscs}
C. Eling and Y. Oz, ``A novel formula for bulk viscosity from the null horizon focusing equation", JHEP \textbf{1106} (2011) 007, [arXiv:1103.1657 [hep-th]].

\bibitem{Buchel:universality_shear}
A. Buchel and J. T. Liu, ``Universality of the shear viscosity in supergravity", Phys. Rev. Lett. \textbf{93} (2004) 090602, [arXiv:hep-th/0311175].

\bibitem{Balasubramanian:stressT_AdS}
V. Balasubramanian and P. Kraus, ``A stress tensor for Anti-de Sitter gravity", Commun. Math. Phys. \textbf{208} (1999) 413, [arXiv:hep-th/9902121].

\bibitem{Kanitscheider:non-confml_holorphy}
I. Kanitscheider, K. Skenderis and M. Taylor, ``Precision holography for nonconformal branes", JHEP \textbf{0809} (2008) 094, [arXiv:0807.3324 [hep-th]].

\bibitem{Bigazzi:SS_dyn_flavr}
F. Bigazzi and A. L. Cotrone, ``Holographic QCD with dynamical flavors", JHEP \textbf{1501} (2015) 104, [arXiv:1410.2443 [hep-th]].


\bibitem{Finazzo:transpt_coeficit_non-conf_QGP}
S. I. Finazzo, R. Rougemont, H. Marrochio and J. Noronha, ``Hydrodynamic transport coefficients for the nonconformal QGP from holography'',  JHEP \textbf{1502} (2015) 051 [arXiv:1412.2968 [hep-ph]].

\bibitem{Wuchao:D0D4_SS}
Chao Wu, Zhiguang Xiao and Da Zhou, ``Sakai-Sugimoto Model in D0-D4 Background",  Phys. Rev. \textbf{D 88} (2013) 026016, [arXiv:1304.2111 [hep-th]].

\bibitem{LiaoJinfeng:anomalous_effect_HIC}
Jinfeng Liao, ``Anomalous effects and possible environmental symmetry ``violation" in heavy-ion collisions", Pramana \textbf{84} (2015) No.5, 901, [arXiv:1401.2500 [hep-ph]].

\end{thebibliography}
\end{document}